\documentclass[12pt]{article}

\usepackage{graphicx}
\usepackage{bm}
\usepackage{array}
\usepackage{tablefootnote}
\usepackage{hyperref}

\begin{document}

\title{Probing a divergent van Hove singularity of graphene with a Ca$_2$N support: a layered electride as a solid-state dopant}

\author{Takeshi Inoshita$^{1,2}$, Masaru Tsukada$^3$, Susumu Saito$^4$ and Hideo Hosono$^1$}


\date{%
    $^1$\textit{\small Materials and Structures Laboratory, Tokyo Institute of Technology, Japan} \\
    $^2$\textit{\small National Institute for Materials Science, Japan}  \\
    $^3$\textit{\small Advanced Institute for Materials Research, Tohoku University, Japan}\\
   $^4$\textit{\small Department of Physics, Tokyo Institute of Technology, Japan}  \\  
    }
    
\maketitle

\begin{abstract}
Layered electrides, as typified by Ca$_2$N, are a new class of quasi-two-dimensional materials with low work functions.   Using first-principles calculations, we have shown that a graphene layer deposited on Ca$_2$N is doped to $n=5\times 10^{14}$ cm$^{-2}$ with its Fermi level aligned with the logarithmically divergent van Hove singularity in the graphene $\pi^*$ band.  For bilayer graphene on Ca$_2$N, the inner graphene layer is doped to the same level, while the doping of the outer graphene layer is much more modest. This finding opens an interesting possibility of using layered electrides for the exploration of van Hove physics.  The work function changes nonmonotonically with the number of graphene layers, which we explain in terms of the peculiar electronic structures of the constituent materials and their bonding.
\end{abstract}

\section{Introduction}
In condensed matter research, it is often witnessed that a material known and studied in chemistry for decades finally attracts the attention of physicists and begins to be investigated with a broader perspective and added momentum.   Such is the current status of the family of compounds called electrides, crystals in which electrons act as anions.\cite{Dye1,Dye2}  From the formal valence viewpoint, electrides possess excess electrons (anionic electrons), which mainly accumulate in the interstitial voids in the crystal lattice (as if they are nuclear-free atoms) rather than being associated with ions or atoms.  Of particular interest are layered electrides with quasi-two-dimensional (Q2D) band structures such as Ca$_{2}$N, in which the excess electrons accumulate in the gap between the ionic layers to form a free-electron-like interlayer band.\cite{Le,Wa,PRX,Tada,PRB}   If this band is the sole energy band crossing the Fermi level, as in Ca$_2$N, Sr$_2$N, and Ba$_2$N, it dominates the low-energy properties, such as the electronic conductivity, of the material.  

Since the first demonstration of Ca$_2$N as a Q2D electride,\cite{Le}  studies on the magnetic,\cite{PRX,PRB,Zhang, Zhang2, Otani, Park} optical,\cite{Guan} and mechanical\cite{He} properties of layered electrides have been reported.  Angle-resolved photoemission spectroscopy (ARPES) measurements have successfully identified the Q2D anionic bands of Ca$_2$N\cite{Oh} and Y$_2$C,\cite{Horiba} which are in excellent agreement with those predicted by ab initio calculations based on density functional theory (DFT).  The surface electronic structures of layered electrides have been calculated recently: it was revealed that they possess an extra-surface state, i.e., a two-dimensional (2D) free-electron-like state with the electrons literally floating outside the surface.\cite{InoSurf}   In other words, anionic electrons exist not only in the bulk but also immediately outside the crystal. 
 
The thinning of layered materials down to a few layers to form genuinely 2D materials has been a popular area of research.  An initial attempt to exfoliate layered electrides utilized Scotch tape,\cite{Le} but a more sophisticated liquid-exfoliation technique has been developed recently,\cite{Druffel} which succeeded in producing few-monolayer samples of Ca$_2$N. With this breakthrough, layered electrides have entered the club of 2D materials.\cite{Nov}   

Turning to applications, the use of layered electrides in electronics, \cite{Le} batteries,\cite{Hou} and catalysis \cite{YJKim, Kitano} and as reducing agents in chemical reactions\cite{YJKim, Red} has been proposed.  Many of these applications utilize the low work functions of layered electrides.\cite{Le, Ming}  However, the low work function of layered electrides enables their more direct application as solid-state dopants.  A recent paper reported electron doping into MoTe$_2$ from a Ca$_2$N support with sufficiently  high concentration to induce a structural phase transition in MoTe$_2$.\cite{MoTe2}

In the present paper, we report our DFT electronic structure calculations for graphene/Ca$_2$N heterointerfaces.  The aims of our study were twofold.  The first was to examine whether the low work function of Ca$_2$N leads to the heavy doping of graphene, particularly focusing on the possibility of aligning the Fermi level to the logarithmically divergent van Hove singulariy (VHS) of the graphene $\pi^*$ band.   This aim was motivated by the various instabilities and unconventional phenomena predicted to occur when $E_F$ is aligned to a VHS.\cite{Fleck, Alvarez, Makogon, Gonzalez, Rice, Lu, Halboth}   Our second aim was to clarify the characteristics of the interfacial electronic structure of graphene/Ca$_2$N with the expectation that the results will shed light on the interfaces between layered electrides and van der Waals crystals in general. 
           
The paper is organized as follows.  Section II describes the geometical model for the interface and the method of calculation based on DFT.   After reviewing the surface electronic structure of Ca$_2$N in Sec.~IIIA, the calculation results for MLG/Ca$_2$N and BLG/Ca$_2$N  (MLG and BLG denote monolayer graphene and bilayer graphene, respectively) are presented in Secs.~IIIB and IIIC, respectively, with emphasis on the doping of graphene.    Finally, the work functions are given in Sec.~IIID, and their dependence on the number of graphene layers is discussed in terms of the peculiar electronic structures of the constituent materials and their mutual bonding.

\section{Methodology}
Ca$_2$N is a layered compound having an anti-CdCl$_2$ structure (space group $R \bar{3} m$) with a primitive unit cell containing three atoms, two Ca and one N.  Its conventional unit cell, containing three primitive unit cells, is hexagonal and comprises nine atomic layers forming the same (albeit staggered) triangular lattice with an in-plane nearest-neighbor distance of 3.64 \AA.\cite{Keve}   The nine layers, in turn, are arranged into three layer units (LUs), each made of three layers (Ca/N/Ca)  stacked with a small distance of 1.23 {\AA}.    The gap between these LUs is much larger  (3.81~{\AA}) and accommodates anionic excess electrons.  Using the standard oxidation numbers of Ca and N  (+2 and -3, respectively), the charge arrangement in the $c$ direction of bulk Ca$_2$N can be written schematically as $\cdots$ [CaNCa]$^{+}\cdot $e$^- \cdot $[CaNCa]$^{+} \cdot $e$^- \cdot$ [CaNCa]$^{+}\cdots$.  The electrons, mainly confined in the inter-LU gaps, form a free-electron-like interlayer band.   If we truncate the crystal at (001) planes, the charge arrangement becomes $(e^-/2)\cdot$ [CaNCa]$^{+}\cdot $e$^- \cdot $[CaNCa]$^{+} \cdot $e$^- \cdot$ [CaNCa]$^{+}\cdot (e^-/2)$.  Note the emergence of $e^-/2$ at both ends, which is required to maintain charge neutrality of the system.   This extra charge is accommodated in the extra-surface state mentioned above. 

We modeled Ca$_2$N by the eighteen-layer (six-LU) (001) slab shown in Fig.~1(a).   Graphene was placed on top of the slab assuming lattice matching between a 2$\times$2 surface cell of Ca$_2$N and a 3$\times$3 supercell of graphene (Fig.~2), which requires the graphene to be compressed by 1.6~\%.  As a result of extending the unit cell, the K (M) point in the original graphene Brillouin zone is folded back to $\bar{\Gamma}$  ($\bar{\mathrm{M}}$).  [This lattice-matching model was adopted as an approximation to the actual structure,  which may be one with a smaller compression of graphene (a commensurate structure with a larger lateral period or an incommensurate structure).]  Concerning the lateral position of graphene, we considered the configurations in which at least one of the Ca atoms on the surface is located immediately below a symmetric site of graphene.  There are two such configurations indicated as A and V in Fig. 2.    In the case of BLG on Ca$_2$N, we placed another graphene layer on A or V assuming Bernal stacking and obtained the three configurations Aa, Ab, and Va in Fig.~2.   The most stable configurations were determined by comparing the calculated total energies.   

The DFT electronic structure calculations were carried out using the Vienna ab initio simulation package (VASP) \cite{Kr,Kr2} based on the projector augmented wave method and a plane-wave basis set.\cite{Bl, Kr3}  The exchange-correlation energy was treated in the generalized gradient approximation (GGA) proposed by Perdew, Burke, and Ernzerhof (PBE).\cite{PBE}   The plane waves were cut off at 700 eV, and integration over the Brillouin zone was performed on a $7\times 7 \times 1$  ($11\times 11 \times 11$)  $\Gamma$-centered grid for the slabs (bulk). Since graphene sheets are held together by weak van der Waals interactions, which cannot be adequately treated within the GGA, we considered these interactions by Grimme's DFT-D3 approximation.\cite{Gri}  The interlayer distance of graphite obtained by this approximation is 3.44 \AA, in reasonably good agreement with the experimental value of 3.36 \AA.\cite{graphite}  To treat the slab structures with plane waves, supercells were constructed by stacking the slabs periodically in the $c$ direction with a gap of at least 20~{\AA}.   Dipole corrections were included to minimize spurious interactions between the periodic slab images.   The structures were relaxed with extreme care because the systems are soft and convergence was slow.   Electronic iteration was deemed convergent when the changes in the total energy per unit cell and the Kohn--Sham energies between steps became less than 10$^{-7}$~eV.  The ionic relaxation was carried out with sufficiently small steps until the maximum force on the atoms became $<$~0.005 eV/{\AA}.  Subsequently, $\sim$~50 additional ionic iteration steps were executed to ensure convergence. To obtain 2D densities of states with high accuracy, we used the analytic quadratic integration method proposed by Harris.\cite{Harris}

\begin{figure}
\includegraphics[width=12 cm]{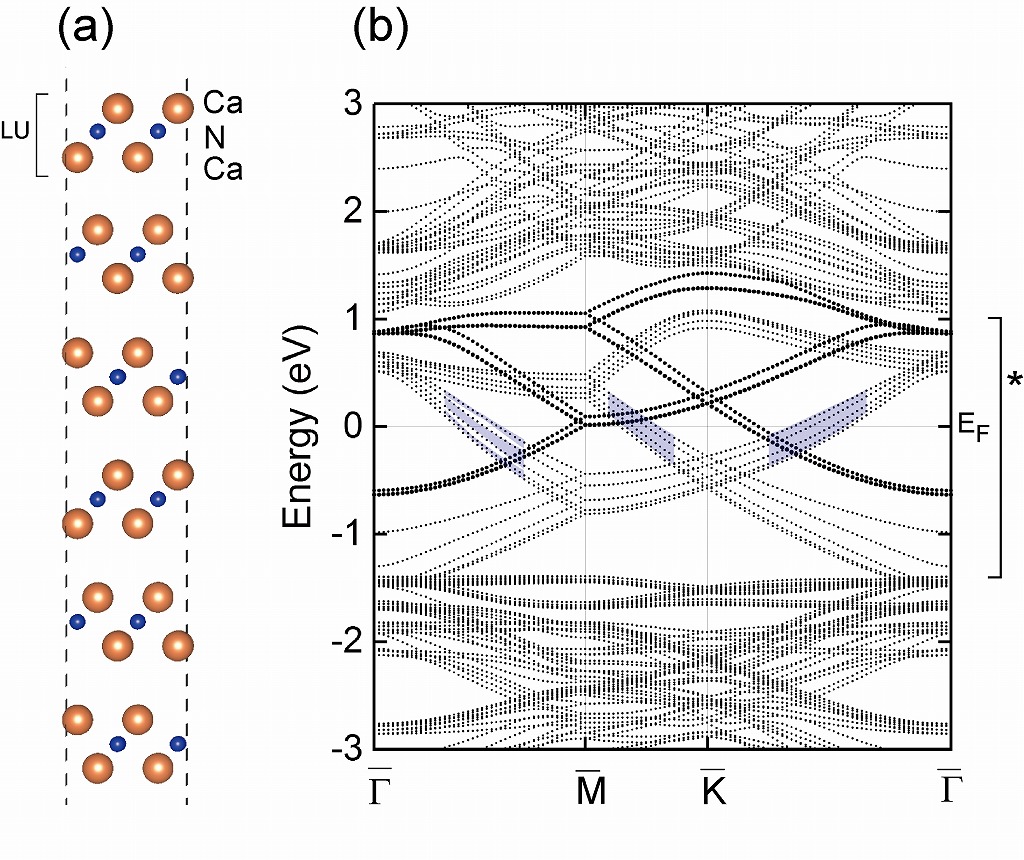}
\caption{\label{}(a) Ca$_2$N (001) slab, consisting of six LUs (18 layers), used in our study. It was optimized by relaxing the upper three LUs while keeping the lower three LUs frozen in the optimized structure for the bulk.   (b) Band structure of the slab in (a) obtained by DFT along the symmetry lines of the 2D hexagonal Brillouin zone.  The bands spanning the energy range between -1.4 and 1~eV are the anionic bands (indicated by the asterisk), which are either bulk states, having charge density mainly confined to the gaps between the LUs, or extra-surface states localized immediately outside the surfaces.   The extra-surface states are highlighted by bold dots.   The light blue shaded areas near the Fermi level indicate the continua of bulk energy bands projected onto the 2D surface Brillouin zone.}
\end{figure}

\begin{figure}
\includegraphics[width=12 cm]{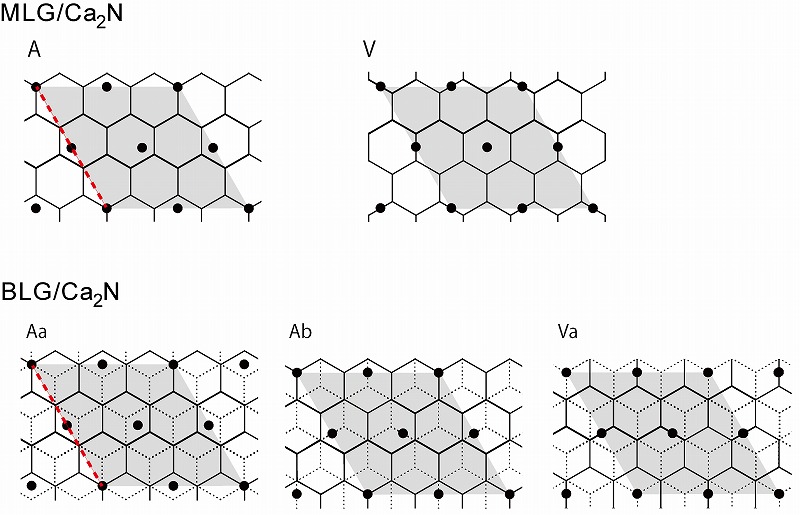}
\caption{\label{}Stacking configurations considered in our calculation.  The shaded areas indicate the unit cell and the spheres denote Ca ions in the top layer of the Ca$_2$N slab.    Configurations A and V: MLG/Ca$_2$N(001) with the lines indicating the hexagonal lattice of graphene.  Configurations Aa, Ab, and Va: BLG/Ca$_2$N(001) with the solid (dashed) lines denoting the lattice of the inner (outer) graphene layer.  Bernal stacking between the two graphene layers is assumed, with Aa and Ab derived from A, and Va from V. The red dashed lines at the left of A and Aa indicate the slice planes used to draw Fig.~4 and Figs.~7(a) and (b).}
\end{figure}

\begin{figure}
\includegraphics[width=9 cm]{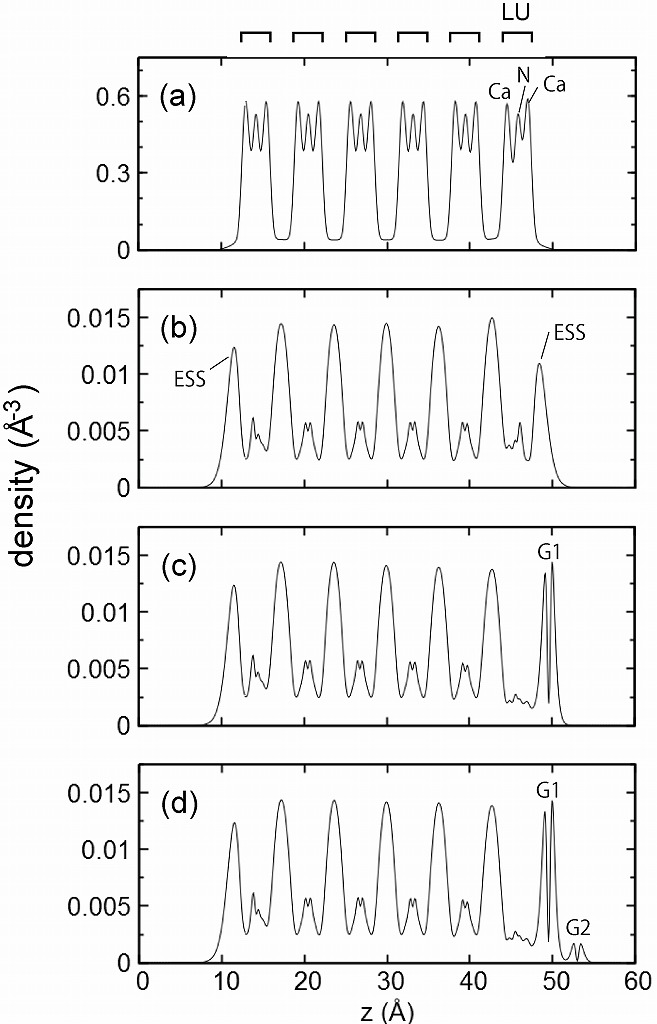}
\caption{\label{}Layer-averaged electron densities for the structures investigated.  (a) Total electron density for the Ca$_2$N slab. The peaks indicate the positions of the ionic layers.  (b) Partial electron density for the Ca$_2$N slab. Note that the dominant peaks are located in the inter-LU gaps and also outside the surface ionic layer (extra-surface states indicated as ESS).  (c) Partial electron density for MLG/Ca$_2$N, where G1 indicates the graphene layer.   (The G1 peak has two subpeaks corresponding to the two lobes of a carbon 2$p_z$ orbital.) (d) Partial electron density for BLG/Ca$_2$N, where G1 and G2 indicate the inner and outer graphene layers, respectively.  (b)-(d) were calculated for electron energies between -0.5 and 0 eV.}
\end{figure}


\begin{figure}
\includegraphics[width=12cm]{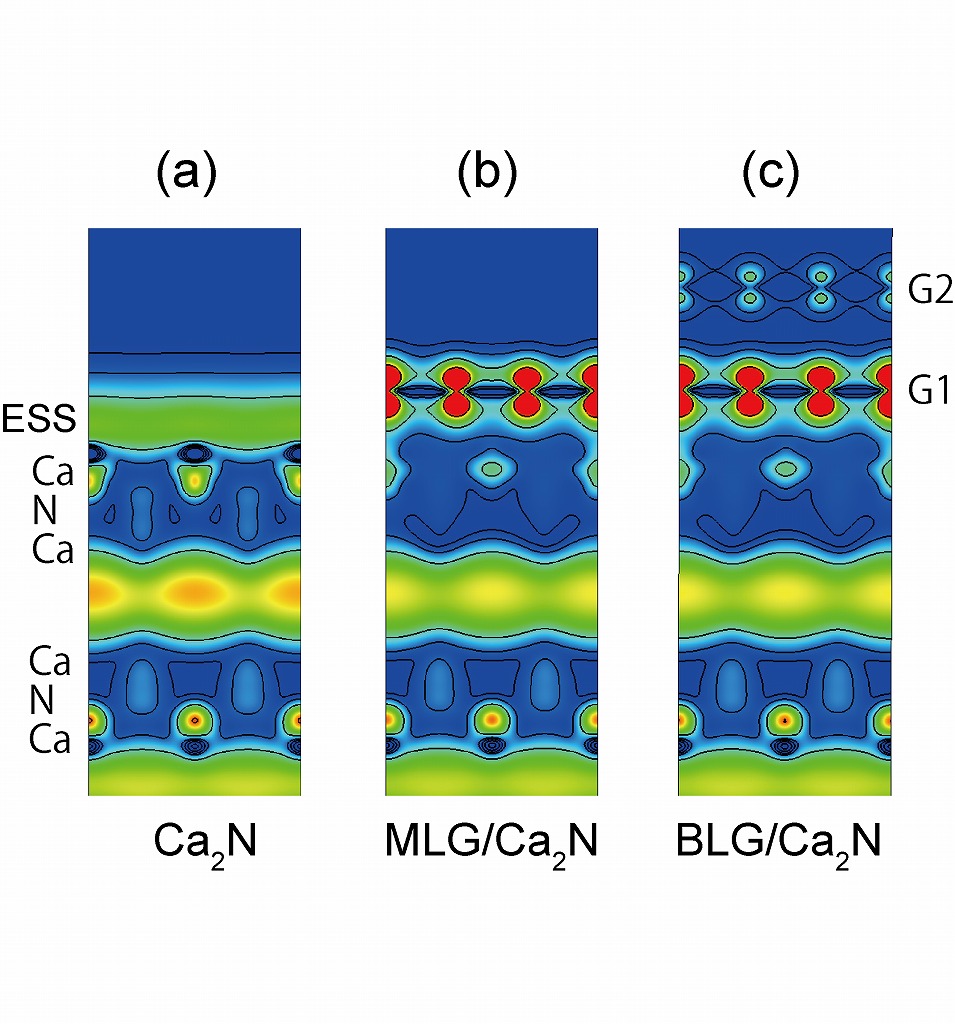}
\caption{\label{}(100) contour maps of the partial charge densities (electron energy between -0.5 and 0 eV) for (a) Ca$_2$N, (b) MLG/Ca$_2$N, and (c) BLG/Ca$_2$N.  The slice plane is that indicated by red dashed lines at the left of A and Aa in Fig.~2.   The color gradient is RGB with an electron density ranging from 0 (blue) to 3$\times 10^{-3}$ (red) bohr$^{-3}$.  The densities on the contour lines are in a ratio of 3 starting from a minimum of $10^{-4}$ bohr$^{-3}$.}
\end{figure}

\begin{figure}
\includegraphics[width=10cm]{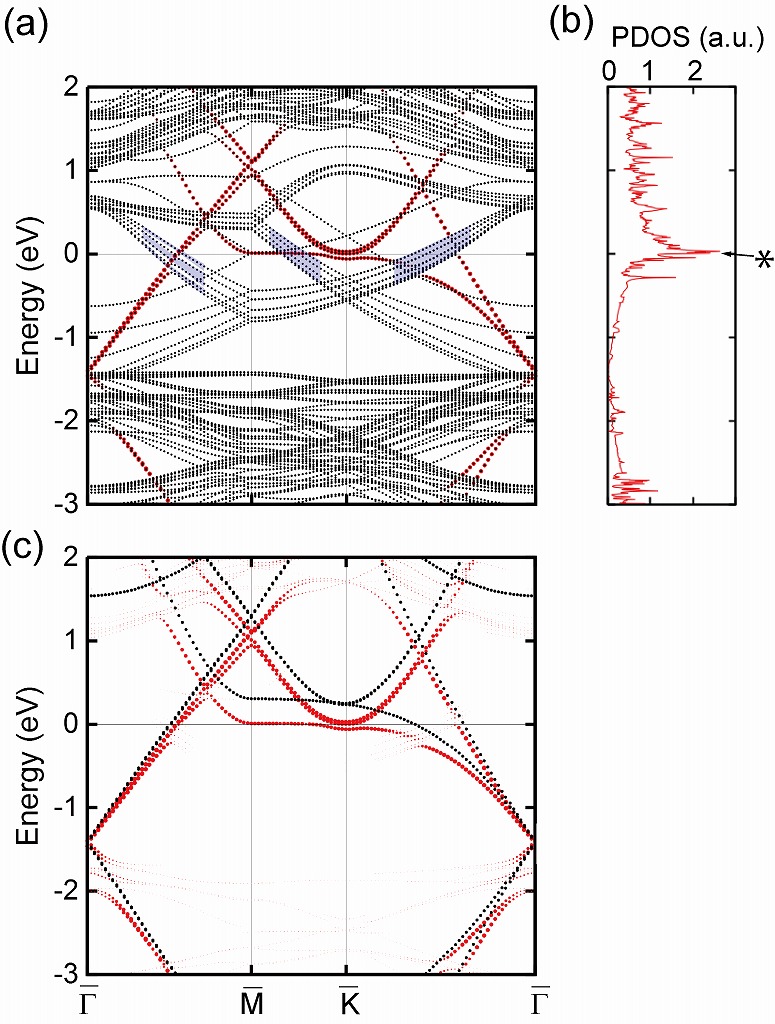}
\caption{\label{}(a) Calculated band structure of MLG/Ca$_2$N.  The $\pi$ bands of graphene are highlighted in red with the radius of each dot proportional to the density of electrons for a given $k$  in the graphene layer. The light blue shaded areas near the Fermi level indicate the bulk energy bands projected onto the 2D surface Brillouin zone (contiuous spectrum at each 2D $k$ point). (b) Partial density of states (PDOS) for graphene calculated from the band structure in (a).  (c) Graphene bands in MLG/Ca$_2$N [red dots in (a)] compared with the bands calculated for an isolated MLG (black dots).  The latter were calculated after removing the Ca and N atoms from the optimized structure of (a) while keeping the positions of the C atoms fixed.  The obtained energy bands were downshifted so that the Dirac cone apex at $\Gamma$ occurs at the same energy as in (a).}
\end{figure}

\begin{figure}
\includegraphics[width=10 cm]{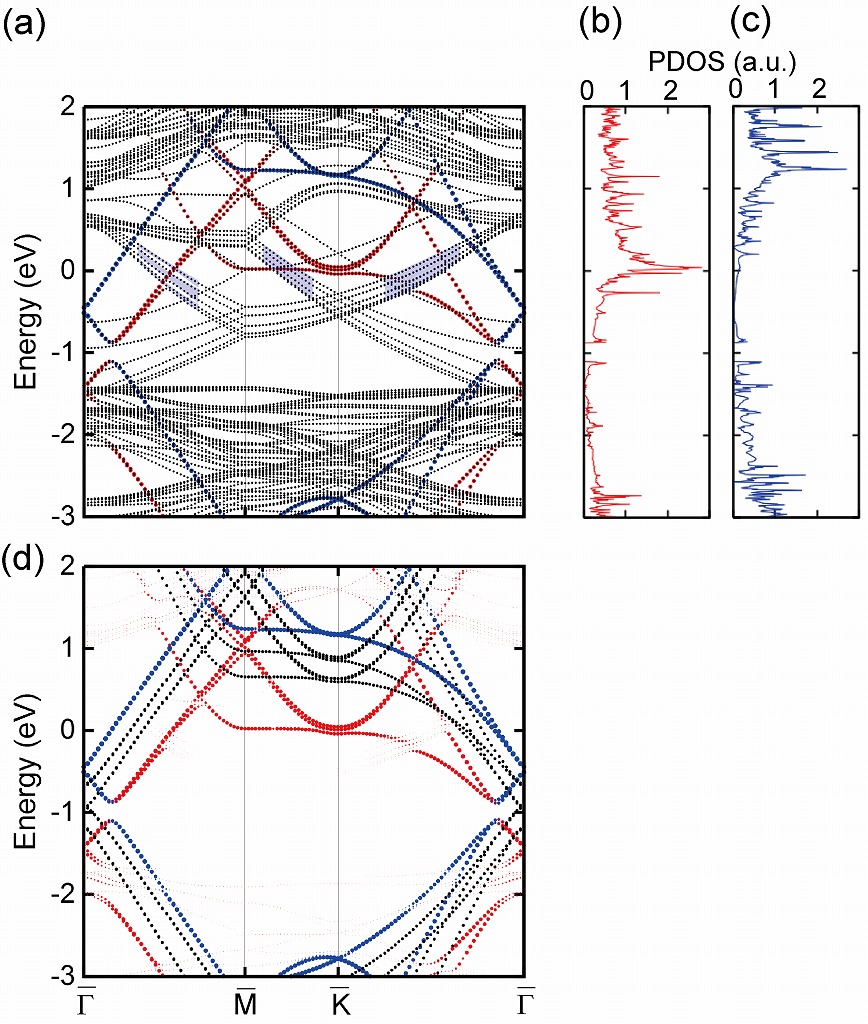}
\caption{\label{}(a) Calculated band structure of BLG/Ca$_2$N.  The $\pi$ bands of the outer graphene layer are highlighted in blue while those of the inner graphene layer (in immediate contact with Ca$_2$N) are shown in red. The radius of each colored dot is proportional to the density of electrons at each $k$ in the corresponding graphene layer. The light blue shaded areas near the Fermi level indicate the bulk energy bands of Ca$_2$N projected onto the 2D surface Brillouin zone.  (b) and (c) Partial density of states (PDOS) for (b) the inner graphene layer and (c) the outer graphene layer calculated from the band structure of (a).  (d) Graphene bands in BLG/Ca$_2$N [red and blue dots in (a)] compared with the bands calculated for an isolated bilayer graphene (black dots).  The latter were obtained after removing the Ca and N atoms from the optimized structure while keeping the positions of the C atoms fixed.  The obtained energies were downshifted to facilitate comparison with (a).}
\end{figure}

\section{Calculation results}
\subsection{Ca$_2$N (001) slab}
We first optimized bulk Ca$_2$N and truncated the obtained structure to obtain a six-LU slab [Fig.~1(a)].   A supercell calculation was then carried out in which the top three LUs were relaxed while keeping the remaining LUs frozen.  Figure 1(b) shows the 2D energy bands calculated for the structurally optimized slab.  The energy bands spanning the energy range between -1.4  and 1~eV, indicated by the asterisk, are anionic bands.  (Throughout this paper, zero energy is taken to be the Fermi energy $E_F$.) Most of these bands are bulk anionic bands with charge density well localized in the inter-LU gaps (interlayer bands). \cite{Le,Wa,PRX} However, there are also anionic bands, highlighted by bold dots, localized outside the surfaces and higher in energy than their bulk counterparts.  These are the {\it extra-surface} bands mentioned in the previous section. Note that these extra-surface bands appear in quasi-degenerate pairs.  In each pair, the upper (lower) band is the surface band associated with the top (bottom) surface of the slab.   They are nondegenerate because we relaxed only the upper surface, and, therefore, the lower and upper surfaces have slightly different structures.  [The nondegeneracy is not due to tunnel coupling between the top and bottom surfaces arising from the finite thickness of the slab.  If both surfaces are relaxed to have an identical structure, the splitting between the surface bands is undetectably small in the energy scale of Fig. 1(b), indicating that our slabs are sufficiently thick to simulate the surface electron states.]

Figure 3 shows the (a) total  and (b) partial layer-averaged electron densities for the Ca$_2$N slab.  The partial electron density accounts for the electrons in the vicinity of the Fermi level (energy between -0.5 and 0 eV).  Note that the peaks in (b) denoted  as ESS lie outside the outermost peaks in (a) arising from the surface Ca layers.  The ESS peaks indicate the extra-surface states characteristic of layered electrides.

\section{Calculation results}
\subsection{Ca$_2$N (001) slab}
We first optimized bulk Ca$_2$N and truncated the obtained structure to obtain a six-LU slab [Fig.~1(a)].   A supercell calculation was then carried out in which the top three LUs were relaxed while keeping the remaining LUs frozen.  Figure 1(b) shows the 2D energy bands calculated for the structurally optimized slab.  The energy bands spanning the energy range between -1.4  and 1~eV, indicated by the asterisk, are anionic bands.  (Throughout this paper, zero energy is taken to be the Fermi energy $E_F$.) Most of these bands are bulk anionic bands with charge density well localized in the inter-LU gaps (interlayer bands).\cite{Le,Wa,PRX}     However, there are also anionic bands, highlighted by bold dots, localized outside the surfaces and higher in energy than their bulk counterparts.  These are the {\it extra-surface} bands mentioned in the previous section. Note that these extra-surface bands appear in quasi-degenerate pairs.  In each pair, the upper (lower) band is the surface band associated with the top (bottom) surface of the slab.   They are nondegenerate because we relaxed only the upper surface, and, therefore, the lower and upper surfaces have slightly different structures.  [The nondegeneracy is not due to tunnel coupling between the top and bottom surfaces arising from the finite thickness of the slab.  If both surfaces are relaxed to have an identical structure, the splitting between the surface bands is undetectably small in the energy scale of Fig. 1(b), indicating that our slabs are sufficiently thick to simulate the surface electron states.]

Figure 3 shows the (a) total  and (b) partial layer-averaged electron densities for the Ca$_2$N slab.  The partial electron density accounts for the electrons in the vicinity of the Fermi level (energy between -0.5 and 0 eV).  Note that the peaks in (b) denoted  as ESS lie outside the outermost peaks in (a) arising from the surface Ca layers.  The ESS peaks indicate the extra-surface states characteristic of layered electrides.   

\subsection{MLG on Ca$_2$N (001)}
For MLG/Ca$_2$N, the calculated total energy for configuration A was found to be lower than that for configuration V by only 0.71~meV per C atom, and the band structures for the two configurations are nearly identical.  We therefore present only the results for A below.  (See Appendix for a comparison between the band structures for the two configurations.) 

The optimum separation $d$ between the graphene and the topmost Ca layer was found to be 2.59~\AA, which indicates that the bonding of graphene to Ca$_2$N is intermediate between chemisorption ($d < 2.3$ {\AA}) and physisorption ($d \sim 3.3$ {\AA}).\cite{Khom}   The flatness of graphene is well preserved with the maximum deviation of the $z$ coordinate of carbon from the average being 0.13 {\AA}.   

The layer-averaged partial electron density (energy between -0.5 and 0 eV) is plotted in Fig.~3(c).   By comparing it with Fig.~3(b), one can see that the extra-surface state peak associated with the upper surface of the Ca$_2$N slab [right ESS peak in Fig.~3(b)] has disappeared and appears to be replaced by that of graphene [G1 in Fig.~3(c)]).   This means that most of the electrons in ESS transfer to graphene.

These findings are corroborated by the contour plots of the partial electron densities for Ca$_2$N and MLG/Ca$_2$N shown in Figs.~4(a) and (b), respectively.  The electrons in ESS accumulated outside the surface of Ca$_2$N are observed to disappear with the deposition of graphene.  The charge densities are affected only down to the gap between the first and second LUs.   

The band structure is shown in Fig.~5(a), where the graphene-derived bands are highlighted in red.  (The radius of each red dot is proportional to the total number of $\pi$ electrons in the atomic sphere of carbon.)   By comparing Fig.~5(a) with Fig.~1(b), it can be readily seen that the energy bands of Ca$_2$N remain largely unchanged except that the extra-surface bands associated with the upper surface  [the upper band of each quasi-degenerate pair of bands represented by bold dots in Fig.~1(b)] have disappeared completely, which is consistent with Figs.~3(b) and (c) and Figs.~4(a) and (b).  The graphene bands are disturbed only where they cross, and hybridize with, the bulk Ca$_2$N bands.  The Dirac cone apex is located at 1.4 eV below $E_F$, and the graphene is heavily $n$-doped with $n=5 \times 10^{14}$/cm$^2$  (0.13 electrons/C).  (To be precise, two Dirac cone apexes overlap here because the apexes at K and K' in the original Brillouin zone of graphene are folded back onto the same $\Gamma$ point.)  The 1.4~eV increase in $E_F$ relative to the Dirac cone apex is significantly higher than those predicted by DFT for various substrates ($ < 0.5$~eV); for example, values of 0.35, 0.49, 0.47, and 0.40 eV have been reported for Ag,\cite{AAC} Al,\cite{AAC} Cu,\cite{AAC} and SiC \cite{SiC} substrates, respectively.  Note, in particular, that $E_F$ is nearly aligned with the 2D VHS  at $\bar{\mathrm{M}}$ (M in the original Brillouin zone of graphene), where the band dispersion has a saddle point causing the density of states to diverge logarithmically.   This singularity is observed clearly in the partial density of states of graphene shown in Fig.~5(b), where the VHS is indicated by an asterisk. Hybridization with Ca$_2$N does not destroy the VHS because there is no bulk band (blue shaded areas in Fig.~5) to hybridize with graphene in the vicinity of $\bar{\mathrm{M}}$.
     
Figure 5(c) shows the graphene bands of MLG/Ca$_2$N plotted in red [the same as the red points in (a)] together with those calculated for isolated MLG plotted in black.   The latter were calculated after removing the Ca and N atoms from the optimized MLG/Ca$_2$N structure.   The energy bands obtained were downshifted rigidly so that the Dirac cone apex occurs at the same energy as that in (a).   (This means that the effect of doping was not taken into account in obtaining the black lines.)  Aside from the difference in band widths and the occurrence of anticrossings due to hybridization with Ca$_2$N, the red and black curves are in good qualitative agreement.   The difference in band widths may be ascribed to the heavy doping of graphene, which brings about band renormalization from enhanced electron-electron interactions.  
   
\subsection{BLG on Ca$_2$N (001)}
For BLG/Ca$_2$N, the total energy of configuration Aa was found to be only slightly lower than those of Va and Ab, the difference being almost negligible (0.65~meV for Va and 0.22~meV for Ab per C atom at the interface, i.e., C atom in the inner graphene layer), and the band structures are nearly indistinguishable (see Appendix). We therefore limit our discussion below to configuration Aa.

The separation between the inner graphene layer and the surface Ca layer is 2.58~{\AA}, almost unchanged from the value of 2.59~{\AA} in MLG/Ca$_2$N, and the distance between the two graphene layers is 3.46~{\AA}, in agreement with the interlayer distance in graphite (3.44~{\AA}) obtained by the same method.  These results are consistent with the small difference between the partial electron density of  MLG/Ca$_2$N [Figs.~3(c) and 4(b)] and that of BLG/Ca$_2$N [Figs.~3(d) and 4(c)]. 

The calculated band structure is shown in Fig.~6(a), where the red (blue) dots denote the bands derived from the inner (outer) graphene layer.   These graphene-like  bands retain much of their original dispersions except where they hybridize with each other (an anticrossing at about -1 eV) and with the bulk (Ca$_2$N) bands.  The band dispersions for the inner graphene layer (red) are nearly identical to those in Fig.~5.  The doping level is also unchanged, with $E_F$ pinned near the VHS of the inner graphene layer, as can be seen from the partial density of states in Fig.~6(b).  The doping of the outer graphene layer (blue) is much more modest, as indicated by the fact that its Dirac cone apex is located only 0.4 eV below $E_F$ [Figs.~6(a) and (c)].   This is consistent with the small peak height of G2 compared with that of G1 in Fig.~3(d) and the low electron density at G2 in Fig.~4(c).  The large difference in doping between the two graphene layers indicates that there is a sizable electric field between them.  
The graphene-derived bands are plotted together with the bands of isolated BLG (black dots) in Fig.~6(d).  (The latter bands were calculated for the structure obtained by removing Ca and N atoms from BLG/Ca$_2$N.  The resulting bands were downshifted to facilitate their comparison with the graphene bands in BLG/Ca$_2$N.)  The graphene bands of BLG/Ca$_2$N can be understood from those of isolated BLG if one takes account of the potential gradient between the two layers and band narrowing due to doping-induced band renormalization (mass enhancement). The fact that the bands derived from the less-doped outer graphene layer are wider than those of the inner graphene layer is consistent with this interpretation.  

\subsection{Work functions}
The work functions $\Phi$ of Ca$_2$N, MLG/Ca$_2$N, and BLG/Ca$_2$N, together with those for MLG and BLG, were calculated by the macroscopic average method.\cite{Fall}  The results are summarized in Table I.      Interestingly, $\Phi$ varies nonmonotonically as a function of the number of graphene layers, decreasing as the first layer is attached and then increasing to a value slightly larger than that of Ca$_2$N as the second layer is deposited.  This trend may be understood by inspecting how the electrons relocate themselves with the deposition of graphene.   

\begin{figure}
\includegraphics[width=8.5cm]{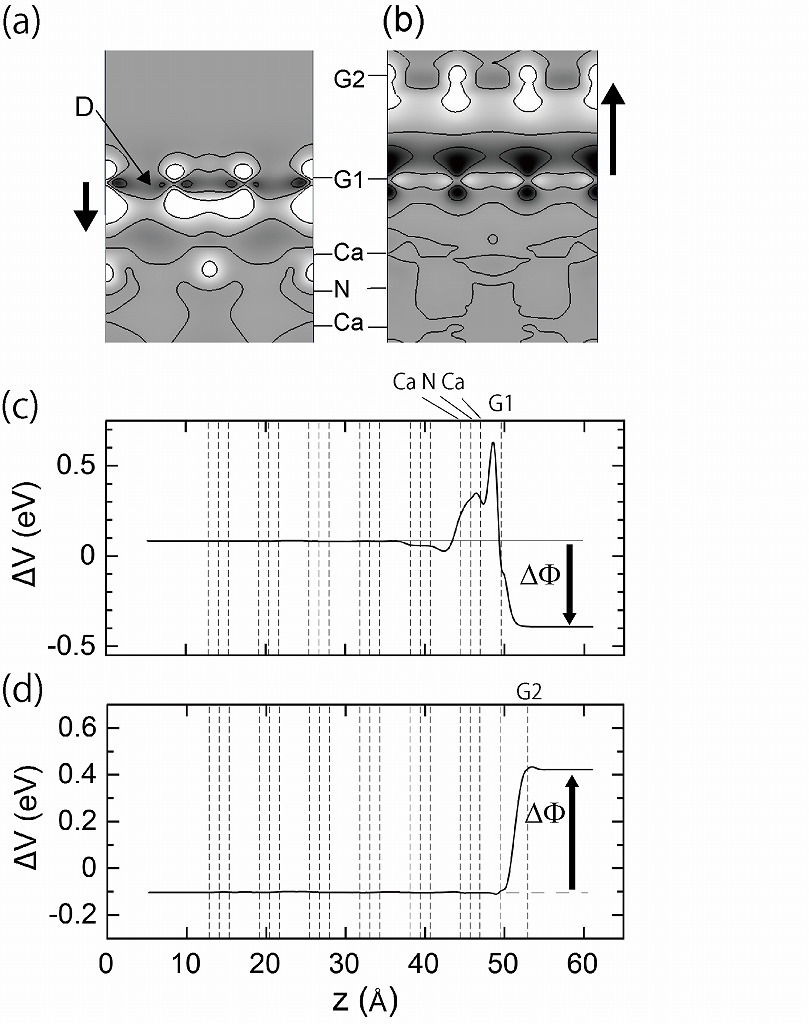}
\caption{\label{} \footnotesize (a) (100) contour map of the difference electron density for MLG/Ca$_2$N defined as the difference between the electron density of the MLG/Ca$_2$N slab and the sum of the electron densities of the graphene layer and the Ca$_2$N slab. D in the figure indicates where the depletion of electrons dominantly takes place.   The gray scale ranges from -5$\times 10^{-3}$ (black)  to 3$\times 10^{-3}$ bohr$^{-3}$ (white),  and the contours are in steps of 2.5$\times 10^{-3}$ bohr$^{-3}$ starting from a minimum of  -5$\times 10^{-3}$. (b) (100) contour map of the difference electron density for BLG/Ca$_2$N defined as the difference between the electron density of the BLG/Ca$_2$N slab and the sum of the electron densities of the graphene layer G2 and the MLG/Ca$_2$N slab. The gray scale ranges from -5$\times 10^{-4}$ (black) to  3$\times 10^{-4}$ bohr$^{-3}$ (white), and the contours are in steps of 2.5$\times 10^{-4}$ bohr$^{-3}$ starting from a minimum of  -5$\times 10^{-4}$ bohr$^{-3}$. The arrows in (a) and (b) indicate the direction of electron transfer as a graphene layer is deposited.    (c) Layer-averaged electrostatic potential $V(z)$ for MLG/Ca$_2$N minus the sum of $V(z)$ for graphene and Ca$_2$N.  (d) Layer-averaged electrostatic potential $V(z)$ for BLG/Ca$_2$N minus the sum of $V(z)$ for graphene and MLG/Ca$_2$N.  In (c) and (d), $\Delta \Phi$ indicates the change in work function due to the attachment of the (last) graphene layer. The positions of the atomic layers are indicated by dashed vertical lines with G1 and G2 denoting the inner and outer graphene layers, respectively.}
\end{figure}

Figure~7(a) shows a (100) map of the difference electron density for MLG/Ca$_2$N defined as the difference between the electron density of MLG/Ca$_2$N  and the sum of the electron densities of the graphene layer and the Ca$_2$N slab.  The map shows electron rearrangement associated with the deposition of graphene in gray scale between white (electron accumulation) and black (depletion).   The electrons are observed to be depleted in the dark horizontal strip (indicated by D) below the graphene layer, reflecting the disappearance of the extra-surface state.   The main destinations of the transferred electrons are the regions in white: these include the lobes around carbon ($n$ doping of graphene), but some electrons also transfer to the bulk, reaching the first LU (as can be seen from the white circular areas between the Ca and N layers).  Note that the lobes around carbon are highly asymmetric, i.e., the lobes above carbon are compact and $p$-like, but those below are more pronounced, extending prominently into Ca$_2$N.   We interpret this enhanced electron density on the bulk side of graphene as arising from bond formation between carbon and Ca$_2$N.  (Such bonding-induced enhancement of the electron density is known to generally take place between a metal substrate and an adatom.\cite{Lang})   Overall, the electrons move inward (from vacuum into the bulk), as shown by the arrow in Fig.~7(a), producing an electric field (dipolar field), which tends to lower the work function. 

The surface contribution to the work function is equal to the change in electrostatic potential across the surface.\cite{Lang2}  Figure~7(c) shows the layer-averaged electrostatic potential $V(z)$ for MLG/Ca$_2$N minus the sum of $V(z)$ for graphene and Ca$_2$N, which we denote by $\Delta V(z)$.  The increase in $\Delta V(z)$ to above zero between graphene and the first LU  (43 \AA$<z<$50 \AA) and its sharp drop to a negative value at graphene (G1) are consistent with the charge rearrangement observed above.

We can analyze the increase in $\Phi$ with the deposition of the second graphene layer (G2) in the same manner.   Figure 7(b) shows a (100) map of the difference electron density for BLG/Ca$_2$N defined as the difference between the electron density of BLG/Ca$_2$N  and the sum of the electron densities of the individual graphene layer (G2) and the MLG/Ca$_2$N slab. The map shows clearly that electron transfer is predominantly from the inner graphene layer (G1) to the outer graphene layer (G2).  This is consistent with the difference electrostatic potential $\Delta V(z)$ presented in Fig.~7(d), showing the confinement of the electric field between the graphene layers.  The fact that the electrons rearrange themeselves only in the intergraphene space may be understood as follows.  As Table I indicates, the Fermi level of an isolated graphene layer is lower than  that of MLG/Ca$_2$N by about 1 eV.  This means that, as the second graphene layer (G2) is attached to the MLG/Ca$_2$N ``substrate'' (G1+Ca$_2$N), electron transfer takes place from the substrate to G2.   This transfer hardly affects the electron distribution in the substrate because its Fermi energy is pinned to the VHS.  Therefore, charge rearrangement associated with the addition of G2 takes place mainly on the vacuum side of the substrate, producing a sharp increase in the electrostatic potential between the two graphene layers and increasing the work function.  With the deposition of further graphene layers, we expect $\Phi$ to increase gradually towards that of graphite.  


\begin{table}
\caption{Work functions for Ca$_2$N(001), MLG/Ca$_2$N(001), and BLG/Ca$_2$N(001) together with those for graphene. The values are taken from the present calculation unless stated otherwise.}
\begin{center}
\begin{tabular}{c|c|c|c|c}  \hline
Ca$_2$N & MLG/Ca$_2$N & BLG/Ca$_2$N & MLG & BLG  \\  \hline
3.39, 3.43$^1$, 3.5$^2$  & 2.95 & 3.47 & 3.94$^3$, 4.26$^4$ & 3.97$^3$, 4.28$^4$ \\  \hline
\end{tabular}
\end{center}
$^1$Ref.~47(calculation) \\
$^2$Ref.~3 (experiment) \\
$^3$Lattice-matched to Ca$_2$N \\
$^4$Free-standing
\end{table}

\section{Concluding remarks}
We have studied the electronic structure of the heterointerface between graphene and the layered electride Ca$_2$N(001) by DFT with particular emphasis on the doping of graphene.  The main findings are as follows:  (1) Upon the deposition of a graphene monolayer, the extra-surface state, a hallmark of layered electrides, vanishes completely, but otherwise the bands of the constituent materials remain reasonably unperturbed and independent.  (2) The electrons originally in the extra-surface state of Ca$_2$N are mostly transferred to graphene, leading to extremely high $n$ doping of graphene with the Fermi level aligned with the logarithmically divergent VHS of graphene's $\pi^*$ band.  (3) A second graphene layer deposited on MLG/Ca$_2$N  is only mildly doped.   (4) The work function decreases with monolayer deposition but increases to a value close to that of Ca$_2$N with bilayer deposition.  This nonmonotonic dependence of the work function on the number of graphene layers reflects the peculiar electronic structures of the constituent materials and their mutual bonding.  

Inspired by the predictions of exotic ground states, interest in the physics of 2D saddle-point VHSs has been growing recently.   As $E_F$ approaches a VHS, various response functions diverge, and small interactions may lead to significant effects.   Instabilities predicted to date in association with VHSs include magnetism (ferromagnetism,\cite{Fleck, Alvarez} antiferromagnetism,\cite{Alvarez} spin density waves\cite{Makogon}), superconductivity (d-wave,\cite{Alvarez} Kohn-Luttinger type\cite{Gonzalez}), and charge density waves.\cite{Rice}   The possibility of modified Friedel oscillations\cite{Lu} and the breaking of lattice symmetry by deformation of the Fermi surface\cite{Halboth} has also been raised.  An attempt at the heavy doping of graphene to explore its VHS using Ca and K ions absorbed on both sides of an MLG on SiC (chemical doping) has been reported recently.\cite{McChesney}   $E_F$ was tuned near the VHS, but none of the instabilities mentioned above were observed.  The doping of graphene with a layered electride support is a promising alternative to chemical doping with the advantage of providing higher-quality samples with much less contamination and disorder.

Finally, the work function engineering of layered electrides through the deposition of graphene (or possibly other 2D van der Waals crystals) may find applications in electronic devices and catalysis.   

\section{Acknowledgment}
T. I. would like to thank Daniel C. Fredrickson for valuable comments on the manuscript.  This work was supported by the MEXT Elements Strategy Initiative to Form Core Research Center and the JST ACCEL Program.  H. H. and S.S. acknowledge support from JSPS KAKENHI Grant No. 17H06153 and JP25107005, respectively.  Crystal structure graphics were generated  by VESTA.\cite{Momma}   

\clearpage

\vspace{ 32 pt}
\noindent
\textbf{\Large Appendix: Comparison between the band structures with different stacking configurations}
\vspace{12 pt}

Figure 8 presents the band structures near $E_F$ calculated for the two stacking configurations A and V of MLG/Ca$_2$N.  Similar plots for BLG/Ca$_2$N with the three stacking configurations Aa, Ab, and Va are shown in Fig.~9.  In either case, the difference between different configurations can be seen to be minimal.

\begin{figure}
\includegraphics[width=10 cm]{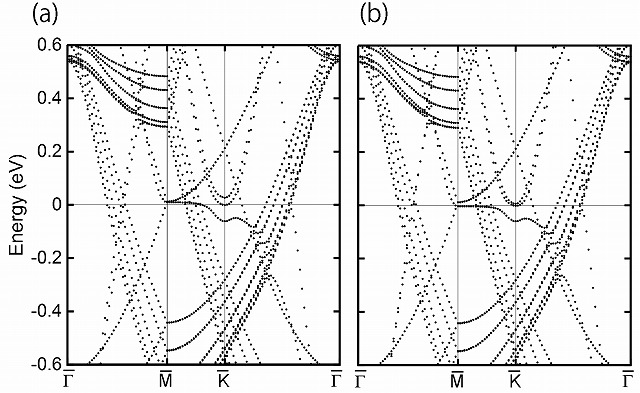}
\caption{\label{}Energy band structures near $E_F$ for MLG/Ca$_2$N with stacking configurations (a) A and (b) V.} 
\end{figure}

\begin{figure}
\includegraphics[width=10 cm]{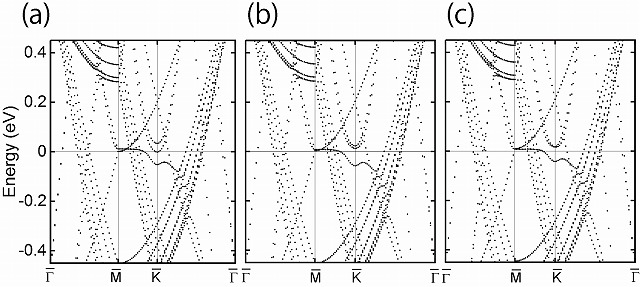}
\caption{\label{}Energy band structures near $E_F$ for BLG/Ca$_2$N with stacking configurations (a) Aa, (b) Ab and (c) Va.} 
\end{figure}


\begin{thebibliography}{}
\bibitem{Dye1} J. L. Dye, Science \textbf{247}, 663 (1990). 
\bibitem{Dye2} J. L. Dye, Science \textbf{301}, 607 (2003).
\bibitem{Le}  K. Lee, S. W. Kim, Y. Toda. S. Matsuishi, and H. Hosono, Nature \textbf{494}, 336 (2013).
\bibitem{Wa} A. Walsh and D. O. Scanlon, J. Mater. Chem. C \textbf{1}, 3525 (2013).
\bibitem{PRX}T. Inoshita, S. Jeong, N. Hamada, and H. Hosono, Phys. Rev. X\textbf{4}, 031023 (2014).
\bibitem{Tada}T. Tada, S. Takemoto, S. Matsuishi, and H. Hosono, Inorg. Chem. \textbf{53}, 10347 (2014).
\bibitem{PRB}T. Inoshita, N. Hamada, and H. Hosono, Phys. Rev. B \textbf{92}, 201109(R) (2015).
\bibitem{Zhang}X. Zhang, Z. Xiao, H. Lei, Y. Toda, S. Matsuishi, T. Kamiya, S. Ueda, and H. Hosono, Chem. Mater. \textbf{22}, 6638 (2014).
\bibitem{Zhang2}X. Zhang, S. Matsuishi, and H. Hosono, J. Appl. Phys. D \textbf{49}, 335002 (2016).
\bibitem{Otani}S. Otani, K. Hirata, Y. Adachi, and N. Ohashi, J. Cryst. Growth \textbf{454}, 15 (2016).
\bibitem{Park}J. Park, K. Lee, S. Y. Lee, C. N. Nandadasa, S. Kim, K. H. Lee, Y. H. Lee, H. Hosono, S-G. Kim, and S. W. Kim, J. Am. Chem. Soc. \textbf{139}, 615 (2017).
\bibitem{Guan}S. Guan, S. A. Yang, L. Zhu, J. Hu, and Y. Yao, Sci. Rep. \textbf{5}, 12285 (2015).
\bibitem{He}Y. He, J. Alloys Compd. \textbf{654}, 180 (2016).
\bibitem{Oh}J. S. Oh, C. J. Kang, Y. J. Kim, S. Sinn, M. Han, Y. J. Chang, B. G. Park, S. W. Kim, B. I. Min, and H. D. Kim, J. Am. Chem. Soc. \textbf{138}, 2496 (2016).
\bibitem{Horiba}K. Horiba, R. Yukawa, T. Mitsuhashi, M. Kitamura, T. Inoshita, N. Hamada, S. Otani, N. Ohashi, S. Maki, J. I. Yamaura, H. Hosono, Y. Murakami, and H. Kumigashira, Phys. Rev. B \textbf{96}, 045101 (2017).
\bibitem{InoSurf}T. Inoshita, S. Takemoto, T. Tada, and H. Hosono, Phys. Rev. B \textbf{95}, 165430 (2017).
\bibitem{Druffel}D. L. Druffel, K. L. Kuntz, A. H. Woomer, F. M. Alcorn, J. Hu, C. L. Donley, and S. C. Warren, J. Am. Chem. Soc. \textbf{138}, 16089 (2016).
\bibitem{Nov}K. S. Novoselov, A. Mishchenko, A. Carvalho, and A.H. Castro Neto, Science \textbf{353},  aac9439 (2016).
\bibitem{Hou}J. Hou, K. Tu, Z. Chen, J. Phys. Chem. C \textbf{120}, 18473 (2016).
\bibitem{YJKim}Y. J. Kim, S. M. Kim, E. J. Cho, H. Hosono, J. W. Yang, and S. W. Kim, Chem. Sci. \textbf{6}, 3577 (2015).
\bibitem{Kitano}M. Kitano, Y. Inoue, H. Ishikawa, K. Yamagata, T. Nakao, T. Tada, S. Matsuishi, T. Yokoyama, M. Hara, and H. Hosono, Chem. Sci. \textbf{7}, 4036 (2016).
\bibitem{Red}Y. J. Kim, S. M. Kim, C. Yu, Y. Yoo, E. J. Cho, J. W. Yang, and S. W. Kim, Langmuir \textbf{33}, 954 (2017).
\bibitem{Ming}W. Ming, M. Yoon, M.-H. Du, K. Lee, and S. W. Kim, J. Am. Chem. Soc. \textbf{138}, 15336 (2016).
\bibitem{MoTe2}S. Kim, S. Song, J. Park, H. S. Yu, S. Cho, D. Kim, J. Baik, D.-H. Choe, K. J. Chang, Y. H. Lee, S. W. Kim, and H. Yang, Nano Lett., \textbf{Article ASAP}, 10.1021 (2017).
\bibitem{Fleck}M. Fleck, A. M. Ole\'{s}, and L. Hedin, Phys. Rev. B \textbf{56}, 3159 (1997).
\bibitem{Alvarez}J. V. Alvarez, J. Gonz\'{a}lez, F. Guinea, and M. A. H. Vozmediano, J. Phys. Soc. Jpn. \textbf{67}, 1868 (1998).
\bibitem{Makogon}D. Makogon, R. van Gelderen, R. Rold\'{a}n, and C. M. Smith, Phys. Rev. B \textbf{84}, 125404 (2011).
\bibitem{Gonzalez}J. Gonz\'{a}lez, Phys. Rev. B \textbf{78}, 205431 (2008).
\bibitem{Rice}T. M. Rice and G. K. Scott, Phys. Rev. Lett. \textbf{35}, 120 (1975).
\bibitem{Lu}C-K. Lu, J. Phys. Condens. Matter \textbf{28}, 065001 (2016).
\bibitem{Halboth}C. J. Halboth and W. Metzner, Phys. Rev. Lett. \textbf{85}, 5162 (2000).
\bibitem{Keve}E. T. Keve and A. C. Skapski, Inorg. Chem. \textbf{7}, 1757 (1968).  
\bibitem{Kr}  G. Kresse and J. Furthm\"{u}ller, Phys. Rev. B \textbf{54}, 11169 (1996).
\bibitem{Kr2}  G. Kresse and J. Furthm\"{u}ller, Comput. Mat. Sci. \textbf{6}, 15 (1996).
\bibitem{Bl} P. E. Bl\"{o}chl, Phys. Rev. B \textbf{50}, 17953 (1994).
\bibitem{Kr3}  G. Kresse and D. Joubert, Phys. Rev. B \textbf{59}, 1758 (1999).
\bibitem{PBE} J. P. Perdew, K. Burke, and M. Ernzerhof, Phys. Rev. Lett. \textbf{77}, 3865 (1996).
\bibitem{Gri}S. Grimme, J. Antony, S. Ehrlich, and S. Kreig, J. Chem. Phys. \textbf{132}, 154104 (2010).
\bibitem{graphite}P. Trucano and R. Chen, Nature \textbf{258}, 136 (1975).  
\bibitem{Harris}F. E. Harris, J. Phys. Condens. Matter \textbf{14}, 621 (2002).
\bibitem{Khom}P. A. Khomyakov, G. Giovannetti, P. C. Rusu, G. Brocks, J. van den Brink, and P. J. Kelly,   Phys. Rev. B \textbf{79}, 195425 (2009).
\bibitem{AAC}C. Gong, G. Lee, B. Shan, E. M. Vogel, R. M. Wallace, and K. Cho, J. Appl. Phys. \textbf{108}, 123711 (2010).    
\bibitem{SiC}F. Varchon, R. Feng, J. Hass, X. Li, B. N. Nguyen, C. Naud, P. Mallet, J.-Y. Veuillen, C. Berger, E. H. Conrad, and L. Magaud, Phys. Rev. Lett. \textbf{99}, 126805 (2007).

\bibitem{Fall}C. J. Fall, N. Binggeli, and A. Baldereschi, J. Phys. Condens. Matter \textbf{11}, 2689 (1999).
\bibitem{Lang}N. D. Lang and A. R. Williams, Phys. Rev. Lett. \textbf{37}, 212 (1976).
\bibitem{Lang2}N. D. Lang, in \textit{Solid State Physics}, vol. 28, ed. H. Ehrenreich, F. Seitz, and D. Turnbull (Academic Press, New York, 1974).
\bibitem{Uij} M. Uijttewaal, G. A. de Wijs, and R. A. de Groot, J. Appl. Phys. \textbf{96}, 1751  (2004).
\bibitem{McChesney}J. L. McChesney, A. Bostwick, T. Ohta, T. Seyller, K. Horn, J. Gonz\'{a}lez, and E. Rotenberg, Phys. Rev. Lett. \textbf{104}, 136803 (2010).
\bibitem{Momma}K. Momma and F. Izumi, J. Appl. Crystallogr. \textbf{44}, 1272 (2011).
\end{thebibliography}
\end{document}